\documentstyle[prl,aps]{revtex}
\newcommand{\beq}{\begin{equation}}
\newcommand{\eeq}{\end{equation}}
\newcommand{\beqn}{\begin{eqnarray}}
\newcommand{\eeqn}{\end{eqnarray}}
\newcommand{\beqa}{\begin{eqnarray}}
\newcommand{\eeqa}{\end{eqnarray}}
\newcommand{\bi}{\bibitem}

\newcommand{\nn}{\nonumber\\}
\newcommand{\bc}{\begin{center}}
\newcommand{\ec}{\end{center}}
\renewcommand{\bf}{\bbox }

\newcommand{\ov}{\overline}
\newcommand{\bx}{{\bbox x}}
\newcommand{\bxp}{{\bbox x'}}

\begin{document}

\title{
\bc
Large time off-equilibrium dynamics of a manifold in a
random potential
\ec}
\author{Leticia F. Cugliandolo}
\address{Service de Physique de L'Etat Condens\'e,
Saclay \cite{add1}, CEA,
Orme des Merisiers, 91191, Gif sur Yvette Cedex, France.
}
\author{Jorge Kurchan and Pierre Le Doussal}
\address{CNRS-Laboratoire de Physique Theorique de
l'Ecole Normale Superieure \cite{add4},
24 rue Lhomond,75231 Cedex 05, Paris, France.
}
\maketitle

%receipt{}
\date{July 12, 1995}
%\begin{document}
%\maketitle
%\widetext
\begin{abstract}
We study the out of equilibrium dynamics
of an elastic manifold in a
random potential using mean-field theory.
We find two asymptotic time regimes:
(i) stationary dynamics, (ii)
slow aging dynamics with violation of equilibrium theorems.
We obtain an analytical solution valid for all large times
with  universal scalings of two-time quantities with  space.
A non-analytic scaling function
crosses over to ultrametricity when the correlations
become long-range.
We propose procedures
to test numerically or experimentally the extent to which
this scenario holds for a given system.

\end{abstract}
%\nopagebreak
%\twocolumn
\narrowtext

%\newpage

The dynamics of an elastic manifold
in a quenched random potential
is relevant for a large number of
experimental systems. Examples are flux lattices
in high-Tc superconductors \cite{revue_russes},
interfaces in random fields
\cite{Naru}, charge density waves \cite{fukuyama_lee_cdw},
dislocation in disordered solids
\cite{vinokur_dislocations}, surface growth on disordered
substrates \cite{Cush}.
The competition between elasticity and disorder
produces a `glass' state that may exhibit pinning,
slow dynamics and non linear macroscopic response
(e.g, leading to zero linear
resistivity in superconductors \cite{feigelman_collective}).
While there is a phenomenological picture
based on scaling arguments (droplets) \cite{revue_russes}
no satisfactory analytical approach is available at present
for the low-temperature dynamics.

The {\it statics} of a $d$-dimensional elastic manifold
embedded in a $N$ dimensional space
in the presence of a random potential
was studied by M\'ezard and Parisi (MP)
who applied a replica variational
Gaussian approximation (Hartree) for $N$ finite which becomes exact
at $N= \infty$  \cite{Mepa}. The replica symmetry breaking (RSB) solution
captures some of the essential physics
in finite $N$ dimension,
such as sample-to-sample susceptibility fluctuations \cite{Mepa,Hwafish},
and predicts non trivial (Flory-like) roughness exponent $\zeta$.
It allows for a theory \cite{Legi} of the statics
of the vortex glass state in superconductors relevant
to experiments such as flux decorations.
Other analytical approaches are based on RG
methods \cite{Legi,xx,Bafi} and it is
as yet unclear whether they capture all
the physics \cite{Legi2,fdtrg}.
Despite the obvious interest of the static approach,
it applies by construction to equilibrium
(Gibbs measure) properties, which may not hold
for experimental times in a glassy system \cite{aging}.

In this Letter we
study, also within the Hartree approximation, the dynamics
of this problem
starting from a random  configuration
as in a temperature quench.
We find that at low enough temperature
there is an  aging
regime and that the system never reaches
equilibrium.
The correlation functions then
depend not only on time differences but also
on the waiting time after the quench.
We obtain the two-time scaling with explicit
space dependence,
a new feature with respect to the mean-field
analytical results for glassy systems obtained so far
\cite{Cuku,Frme,Bacukupa,Cule}. Details
will be presented elsewhere \cite{Cukule}.

This treatment is exact for $N=\infty$.
At finite $N$ it holds {\it a priori} only within the
Gaussian variational ansatz. Whether
the properties  hold qualitatively for true finite $N$ dimensional models
cannot be answered analytically at present.
The good qualitative agreement\cite{Cuku} of the MF dynamical
analytical solution with experiments on
spin glasses suggests that our results for $N=\infty$
may be relevant for some systems related to the present
model.

The main purpose of this Letter is to suggest, on the
basis of the
exact solution for infinite $N$, {\it definite} predictions
for the off-equilibrium
dynamics which can
be checked in numerical simulations and experiments.
Our results provide a basis for a finite-$N$ analysis.
Remarkably, despite the system being out of equilibrium,
some of the results of the replica calculation by MP,e.g $\zeta$,
are shown to carry through to the dynamics, albeit with a different
interpretation in terms of directly observable time-dependent
physical quantities.

The model of a manifold of internal dimension $d$
embedded in a random medium of dimension $N$ is described, in terms of
an $N$ component displacement field ${\bbox \phi}$, ${\bbox \phi} =
( \phi_1, \phi_2,\dots,\phi_N)$,
by the Hamiltonian \cite{Mepa}
\beqa
\label{RM0}
H = \int d^d x\;
\left[ \;
\frac{1}{2} \; (\nabla {\bbox \phi}(\bx ))^2 + V({\bbox \phi}(\bx ),\bx ) +
\frac{\mu}{2} \; {\bbox \phi}^2
\; \right]
\; ,
\nonumber
%\label{model1}
\eeqa
$\mu$ is a mass, which effectively constraints the manifold
to fluctuate in a restricted volume of the embedding space.
$V$ is a Gaussian random potential with correlations
\beqa
\overline{ V({\bbox \phi},\bx ) V({\bbox \phi'},\bxp) } = -
 N \delta^d(\bx -\bxp)
\; {\cal V} \left( ({\bbox \phi}-{\bbox \phi'})^2/N\right)
\; .
\nonumber
\eeqa

We consider the
Langevin dynamics
\beqa
\frac{\partial {\bbox \phi}(\bx ,t) }{\partial t} =
 - \frac{ \delta H}{\delta {\bbox \phi}(\bx ,t)} + {\bbox \eta}(\bx ,t)
\nonumber
\eeqa
with $\langle \eta_\alpha(\bx ,t) \eta_\beta(\bxp,t') \rangle
= 2 T \; \delta_{\alpha\beta} \; \delta^d(\bx -\bxp) \delta(t-t')$.
We let the system evolve from a
spatially translationally-invariant (STI) configuration at $t=0$.
The system remains STI at subsequent times.
We study the dynamics exactly for $N \to \infty$ and, for $N$ finite,
within the dynamical
Hartree approximation which amounts to substituting
${\cal V}$ by an effective $\hat{{\cal V}}$ \cite{Mepa,Cule}.
The `equilibrium' dynamics ({\it \`a la} Sompolinsky \cite{Sozi})
was studied by Kinzelbach and Horner  for general $d$ \cite{Kiho1}.
The case $d=0$ has been studied in Ref. \cite{Frme}
and an analytical solution at large time was given in Ref. \cite{Cule}.

The quantities of interest in the large-time off - equilibrium dynamics
are the two-time correlation
$
C_{xx'}(t,t')$
$=$
$1/N \overline{ \langle {\bf \phi}(x,t) \cdot  {\bf \phi}(x',t') \rangle }
$
and the response
$
R_{xx'}(t,t')$
$=$
$1/N \overline{
\delta \langle {\bf \phi}(x,t) \rangle / \delta {\bf f}(x',t') }|_{{\bf f}=0}
$
where ${\bf f}(x',t')$ is a small perturbation applied at the space point $x'$
at time $t'$. We also define
the mean squared displacement
$
D_{xx'}(t,t')$
$=$
$1/N \overline{\langle ({\bf \phi}(x,t)-{\bf \phi}(x',t'))^2 \rangle}
$
and the correlation
$B_{xx'}(t,t')$
$=$
$1/N \overline{\langle
({\bf \phi}(x,t)-{\bf \phi}(x,t'))
({\bf \phi}(x',t)-{\bf \phi}(x',t')) \rangle}$.
The brackets and overline represent the average over
the thermal noise and the quenched disorder, respectively.
The Fourier-transform $B_k$ and $R_k$
w.r.t. the space difference $x-x'$
are used below, and we denote with tilde equal space
($x=0$) two-time functions,
$\tilde{B}=\int_k B_k$.

A common choice is
$
{\cal V}(z) = (\theta + z)^{1-\gamma} / (2 (1-\gamma))
$.
`Short-range' (SR) correlations correspond to $\gamma (1-d/2) > 1$ and
`long-range'
(LR) correlations
to  $\gamma (1-d/2) < 1$. The static solution \cite{Mepa}
is characterized by two exponents:
in the LR case, an ultrametric ansatz gives
$D_x^{st}=\overline{(\phi(x)-\phi(0))^2} \sim x^{2 \zeta}$
with a roughness exponent $\zeta_{LR}=(2-d/2)/(1+\gamma)$
and a free-energy fluctuation exponent
$\theta$. In the SR case a
one step RSB ansatz gives
$\zeta_{SR}=(2-d)/2$.
Statistical rotational symmetry imposes in
general $\theta = 2 \zeta + d -2$.
The $d=2$ Sine Gordon model is marginal and
solved \cite{Legi} by a one step RSB.

{\it Aging.}
Let us describe the picture that emerges for large times and $\mu >0$.
Equal-time quantities reach their asymptotic
values which do not necessarily coincide with the equilibrium ones
\cite{Cuku,Cukule}.
For two-time quantities we consider two different regimes of the times.

(i)
After a large waiting-time $t_w$, $B_k(\tau+t_w,t_w)$ first grows
with $\tau$ in a manner
independent of $t_w$, from $0$ up to the Edwards-Anderson parameter
for the mode $k$ defined as
$b_k^1 \equiv \lim_{\tau \to \infty} \lim_{t_w \to \infty} B_k(\tau+t_w,t_w)$.
In this time regime the displacement
is  time-translation invariant (TTI); we denote it
$b^F_k(\tau) = \lim_{t_w \to \infty} B_k(\tau+t_w,t_w)$.
The response $r^F_k(\tau) = \lim_{t_w \to \infty} R_k(\tau+t_w,t_w)$ satisfies
fluctuation dissipation (FDT)
$r_k^F(\tau)
= 1/(2 T) \; \partial_\tau b_k^F(\tau)
\;  \theta(\tau)$. The `FDT-regime'
is very much like an equilibration in a state,
the manifold looks pinned with an effective mass $\ov M$,
since $b_k^1=(2T)/(k^2 + \ov M)$ (for $\mu \to 0$).

(ii)
However,
for all $t_w$ and sufficiently large $\tau$, $B_k(\tau+t_w,t_w)$ continues to
grow
beyond $b_k^1$ up to
$b_k^o \equiv \lim_{\tau \to \infty} B_k(\tau+t_w,t_w)$.
The growth of $B_k$ now depends on $t_w$:
the larger
$t_w$ the slower the motion of the system, it {\it ages}.
Thus in this model pinning is a gradual process,
the older the system the more pinned it is but it is not pinned forever.
The aging time-regime is thus defined (for $t_w \to \infty$) as
the times $\tau$ such that  $B_k(\tau+t_w,t_w) > b_k^1$.  We denote
by $b_k(\tau+t_w,t_w)$ and $r_k(\tau+t_w,t_w)$ the displacement and response
in the aging regime, where both TTI and FDT are violated.

As regards the measurements of noise and susceptibility
(i) corresponds to high frequencies while (ii) corresponds
to low frequencies (scaling with the waiting time) such that noise
and susceptibility
depend on $t_w$.
In a domain-growth process  (i) corresponds to the fast
thermal fluctuations
of the spins around their mean magnetisation while  (ii)
corresponds to the actual growth of the domains.
An important measurable quantity is
the susceptibility
$\chi_k(t,t') = \int_{t'}^{t}  ds R_k(t,s) $
of the mode $k$, i.e the total linear response
to an external force of spatial modulation $k$
applied during the interval $[t',t]$.
The results below imply
$\chi_k(t_w + \tau,t_w) = k^{-2} F[k^2 \chi_0(t_w + \tau,t_w)]$
where
$\chi_0(t_w + \tau,t_w)^{-1}$ is a `running' effective mass
which exclusively depends on times through
the local displacement, $\chi_0=\chi_0[\tilde b]$,
and goes to zero at large time separation
$\tau$. The typical internal distance
$\sim \sqrt{\chi_0} \sim \tilde b^{1/\zeta}$
grows slower with $\tau$ as the age $t_w$ increases.

We study $\mu \to 0$ {\it after} the large-times
limit. If one takes $\mu=0$ from the
start, one must take into account diffusion:
$B_{xx}(t,0) \to \infty$ at large $t$ since
$\int dk b^o_k = \infty$. In addition to the
{\it aging} regime defined as $t$, $t' \to \infty$ with $B_{xx}(t,t')$ fixed,
there is then a {\it diffusion} regime where $B_{xx}(t,t')/B_{xx}(t,0)$ is
fixed
\cite{Cukule}.
Our results for $B_k$ and $R_k$
are expected to hold also at $\mu=0$ in the aging regime.

{\it Generalization of equilibrium theorems, two-time scalings.}
Let us describe in more detail the aging regime
as derived below.
For large times, the precise manner in which  TTI is
violated is described by `triangle' relations \cite{Cuku}
involving any three times:
\beq
B_k(t_{min},t_{max})= f_k (B_k(t_{min},t_{int}),B_k(t_{int},t_{max}))
\label{triangle}
\eeq
while the violation of FDT is given by:
\beq
R_k(t,t') =  X_k[B_k(t,t')] \; \partial_{t'} B_k(t,t')
\label{Xeq}
\eeq
where (\ref{Xeq}) means that $X_k$ depends on the
times only through $B_k(t,t')$.
In order to complete this ansatz, we have to specify how $f_k$
and $X_k$ for
different $k$ are related to one another.
This is done below in an algebraic manner,
implying no new hypotheses with respect to the $d=0$ case
\cite{Cuku,Cule}. We find that
\beq
B_k(t,t')=B_k[k,B_o(t,t')]
\label{calor}
\eeq
i.e. all $k$-modes depend on times
{\em only through} the dependence
of one of them. Thus one can use any
two-time function, e.g $B_o(t,t')$
or $\tilde{B}(t,t')$ to parametrise the two-time
dependence. We also show:
\beq
X_k[B_k]=X_k[B_k[k,B_o]]=X_o [B_o]=\tilde{X}[\tilde{B}]
\label{uuuf}
\eeq
i.e. the values of all $X_k$  are the same
for times such that $B_o$ takes the same
value. We have also defined
$\tilde{X}[\tilde{b}(t,t')] = \tilde{r}(t,t')/\partial_{t'}\tilde{b}(t,t')$.
Eq. (\ref{uuuf}) implies that if $X$ is time independent
in the aging
regime, then it takes the same value for all $k$.
Hence, for the  $k$-mode susceptibility we have
$\chi_k(t,t')=\chi_k[k,\chi_0(t,t')]$.
These functional dependences are testable
using two-time parametric plots.
Their explicit forms for this model
are determined below.

We find that: (i) in the FDT-regime
$B_k < b_k^1$, $X_k[B_k]=-1/(2T)$
(ii) in the aging-regime $b_k^0 > b_k  > b_k^1$
there are two distinct cases.
For SR correlations $X_k(b_k) = X$ and $f_k$ has the form
$f_k(u,v)=\jmath_k^{-1}(\jmath_k(u) \jmath_k(v))$.
This implies
$B_k(t,t') = \jmath_k^{-1}(h(t')/h(t))$
where $h(t)$ is increasing and independent of $k$.
For LR correlations $X_k(b_k)$ is a non-constant function of $b_k$.
The function $f_k$ is $f_k(u,v) = \max(u,v)$.

{\it Formal solution for the spatial scaling.}
In order to extract
the scaling properties of this model and
to justify the above ansatz,
we  encode the correlation and
response functions in the superspace order-parameter \cite{Ku}.
At the saddle point, using causality, ${\bbox Q}_{x x'}(1,2)$ reads
\beqa
{\bbox Q}_{x x'}(1,2)
=
C_{x x'}(t_1,t_2) +
(\ov \theta_2 - \ov \theta_1) \;
\left[
\theta_2 \,
\;R_{x x'}(t_1,t_2)
+ \theta_1 \, \; R_{x' x}(t_2,t_1)
\right]
\label{Q12}
\eeqa
The $\theta$'s are Grassmann variables and
we  denote $1 \equiv (t_1,\theta_1, \ov \theta_1)$,
$d1 \equiv dt_1 \, d\theta_1 \, d \ov \theta_1$,
${\bbox \delta}(1-2) \equiv (\theta_2 - \theta_1) \,
(\ov \theta_2 - \ov \theta_1) \, \delta(t_2-t_1)$
 and
$
D^{(2)}(1) \equiv  \partial_{\theta_1} \,
\left(
\partial_{\bar \theta_1} -
\theta_1
\partial_{t_1}
\right)
$.
We use two types of functions of the super-order parameter:
`operator' as in
${\bbox Q}^2 _{x x'}(1,2)$
$=$
$ \int d3 \; {\bbox
Q}_{x x'}(1,3){\bbox Q}_{x x'}(3,2)$
and `pointwise' as in ${\bbox Q}^{\bullet \; 2} _{x x'}(1,2)$
$ =$
$ [{\bbox Q} _{x x'}(1,2)]^2$.

It is now easy to write the equation
of motion for the order parameters, exact
at the mean-field level:
\beqa
\left(
D^{(2)}-\nabla^2+\mu +
\int
d3 {\cal V}^{' \bullet}({\bbox {B}}_{x x}(1,3))
\right)
{\bbox Q}_{x x'}(1,2)
-
 \delta^d(x-x') {\bbox \delta}(1-2)
-
2 \,
\left[
{\cal V}^{' \bullet}({\bbox {B}}_{x x}) {\bbox Q}_{x x'}
\right](1,2)
&=&
0
\label{motion}
\eeqa
where $
{\bbox B}_{x x'}(1,2)=
{\bbox Q_{x x'}}(1,1)+{\bbox Q}_{x x'}(2,2)-2{\bbox Q}_{x x'}(1,2) $.

Using space translational-invariance one finds that all fourier modes
${\bbox Q}_k$ can be expressed in terms of the zero mode
${\bbox Q}_{k=0}\equiv
 {\bbox Q}_o $ through the operator relation:
\beq
{\bbox Q}_k(1,2)
=
\left[
k^2
{\bbox \delta}
+
{\bbox Q}_o^{-1} \right]^{-1} (1,2)
\; .
\label{rel}
\eeq

This is (in an encoded notation) a scaling relation
for the correlation and response
functions involving the two times and spatial separation.
Substituting Eq.(\ref{rel}) in Eq.(\ref{motion}) one can
write a separate equation for each mode.
Hence, we have mapped the problem
of each $B_k(t,t')$, $R_k(t,t')$ into
an effective $d=0$ problem
with a complicated memory kernel
\cite{Levi}.
Up to now we have made no approximations. A  first important result is that
one can now solve numerically the equation for one of the modes
and then recover the space dependence algebraicly  from Eq.(\ref{rel}).
The above property implies that an ansatz as in problems without
space-dependence
\cite{Cuku,Cule} applies
to each mode $B_k(t,t')$, $R_k(t,t')$
and justifies
Eqs. (\ref{triangle}),(\ref{Xeq}).

Now, in the large time-limit any operator function
$F[ {\bbox Q}_A]$
of an order parameter ${\bbox Q}_A$
which can be parametrized by
$f_A,X_A$, yields a new
order parameter ${\bbox Q}_B=F[ {\bbox Q}_A]$ which can also,
at long times, be
parametrized in the same form with $f_B,X_B$.
The explicit computation of $f_B,X_B$ in terms of
$f_A,X_A$ has been done in Ref. \cite{Bacukupa} and when
applied to Eq. (\ref{rel}) yields
explicit functional relations between
${\bbox Q}_k$, ${\bbox B}_k$ and ${\bbox Q}_0$
and, in particular,  Eqs. (\ref{calor}),(\ref{uuuf})
for the components $B_k$, $R_k$.

One can extend this calculation to $O(N)$ spherically constrained
models by letting $\mu$ be a function of time and by imposing
${\bbox Q}_{xx}(1,1)=1$ for all $t_1$.

{\it Explicit calculations.}
We can now establish the large time equations in both regimes.
In the FDT-regime
$r_k^F = - x_F \partial_\tau b_k^F(\tau) $ with  $x_F = -1/(2T)$ and one
finds a single equation:
\beqa \label{fdt}
\frac{d b_k^F(\tau)}{d \tau}
&=&
2 T - (k^2  + \ov{M} + 4 x_F {\cal V}'(\tilde b^1)) b_k^F(\tau)
+ 4 x_F \frac{d}{d \tau}
\int_0^{\tau} d \tau'
{\cal V}'(\tilde{b}^F(\tau-\tau')) b_k^F(\tau')
\eeqa
Neglecting the time derivatives in the
l.h.s. of the full dynamical equations
and integrating over the FDT regime (see \cite{Cuku,Cule})
one finds the equations for  the aging regime:
\beqa
0 &=&
r_k(t,t') \; (k^2+ \overline M)
+   \frac{2 b^1_k}{T} \; {\cal V}''(\tilde b(t,t')) \;
      \tilde r(t,t')
 + 4 \int_{t'}^{t} ds \; {\cal V}''(\tilde b(t,s)) \; \tilde r(t,s) \;
                         r_k(s,t')
\label{reqlt1}
\\
0 &=&
- b_k(t,t') \; (k^2+ \overline M)
+  \frac{2 b^1_k}{T}
\left( {\cal V}'(\tilde b^1)-{\cal V}'(\tilde b(t,t') \right)
+
4 \int_{0}^{t} ds \; \left(
{\cal V}'(\tilde b(t,s)) \; r_k(t,s)
+ {\cal V}''(\tilde b(t,s)) \; \tilde r(t,s) \;
b_k(t,s) \right)
\nn
& &
+ 2T
-
4 \int_{0}^{t'} ds \; \left( {\cal V}'(\tilde b(t,s)) \; r_k(t',s)
+ {\cal V}''(\tilde b(t,s))
\; \tilde r(t,s) \; b_k(t',s) \right)
-
4 \int_{t'}^t ds \; {\cal V}''(\tilde b(t,s)) \; \tilde r(t,s) \;
b_k(s,t')
\;  .
\label{beqlt1}
\eeqa
$\overline M \equiv -4 \lim_{t \to \infty} \int_0^{t} ds \;
{\cal V}''(\tilde b(t,s)) \; \tilde r(t,s)$ is the `anomaly'.
These equations have time-reparametrization invariance
which prevent us from determining $h(t)$.
The `quasi-static' values $\tilde b^1$, $b^1_k$
follow from letting $t' \to t_{-}$ in Eq.(\ref{beqlt1}):
\beq
b^1_k = 2T/(k^2+ \overline M)
\label{beqltt}
\eeq
Similarly, Eq.(\ref{reqlt1}) integrated over $k$, yields either
the high-T solution $\tilde r(t,t_-)=0$
or the low-T condition
\beq
1 =
- 4 {\cal V}''(\tilde b^1) \,
 \int_k \; (k^2+ \overline M)^{-2}
\; .
\eeq
This implies
$\tilde{b}^1 = -4T c_d/(d-2)(-4 c_d {\cal V}''(\tilde{b}^1))^{(d-2)/(4-d)}$
and
$\overline M = (-4 c_d {\cal V}''(\tilde{b}^1))^{2/(4-d)}$
for $d<4$, where
$c_d = \int_k (k^2+1)^{-2}$.
Thus it is not necessary to know the details of the aging solution
to determine $b_k^1$, $\tilde b^1$ and $\overline M$.

{\it FDT regime.}
Defining $\phi(\tau)
= 4({\cal V}'(\tilde{b}^{F}(\tau)) - {\cal V}'(\tilde{b^1}))$,
the Laplace transform w.r.t. $\tau$
of Eq. (\ref{fdt}) yields:
\beqa
b_k^F(s)
&=&
 ( x_F s)^{-1} ( k^2 + \ov M + s - x_F s \phi(s) )^{-1}
\eeqa
At small $\tau$, $b_k^F(\tau) \sim (1-\exp(-A_k \tau))/(x A_k)$
with $A_k = k^2 + \ov M - x_F \phi(0)$. When
$b_k^{int} \sim 1/(x A_k)$ there is a crossover to
a slower regime where one can neglect
the term $s$ ($d/d\tau$)
and find a power law behaviour:
$b_k^F(\tau) = b_k^1 - c (b_k^1)^2 \, \tau^{-\beta}$
with $\beta$ determined by \cite{Kiho1}:
\beq
\Gamma[1-2 \beta] \Gamma[1-\beta]^{-2}
=
x_F Y(\tilde b^1)
\eeq
and $Y(\tilde b^1) = 4 {\cal V}''(\tilde b^1)^2
{\cal V}'''(\tilde b^1)^{-1}
\partial_{\ov M } \ln \int_k (k^2 + \ov M)^{-2}$.
An explicit calculation gives $Y(\tilde b^1) = 4/\tilde{X}(\tilde b^1)$
in terms of the function $\tilde{X}(\tilde b)$ defined
by Eq. (\ref{tildeX}).

{\it Aging regime in short range models.}
Power law models are short range for $\gamma> \gamma_c=2/(2-d)$ and $d \leq 2$.
With the ansatz $X_k[b_k(t,t')] = X$ Eqs.(\ref{reqlt1})
and (\ref{beqlt1}) reduce to a single equation for $b_k(t,t')$.
One must have $X=-\ov M/(4 {\cal V}'(\tilde b^1))$.
As discussed above $b_k(t,t') = j_k^{-1}(h(t')/h(t))$.
Defining $u = \ln h(t)$ one has
$b_k(t,t') = {\cal B}_k(u-u')$ where $0<u<\infty$ and
${\cal B}_k(0) = b_k^1$ and ${\cal B}_k(\infty) = b_k^0$.
We obtain:
\beqa
0 =
2 T - (k^2 + \ov{M} + 4 X {\cal V}'(\tilde b^1)) {\cal B}_k(u)
+ 4 X \frac{d}{du}
\int_0^{u} du'
{\cal V}'(\tilde{\cal B}(u-u')) {\cal B}_k(u')
+ 4 (X-x_F)
({\cal V}'(\tilde b^1) - {\cal V}'(\tilde{\cal B}(u)) )
\nonumber
\eeqa
Remarkably, this equation is formally similar to Eq. (\ref{fdt})
though in a completely different variable. Laplace transforming
w.r.t. $u$ one gets:
\beqa
{\cal B}_k(s) = b_k^1 + (X s)^{-1}
[( k^2 + \ov M )^{-1} - ( k^2 + \ov M  - X s \phi(s) )^{-1}]
\nonumber
\eeqa
with $\phi(u) = 4 ({\cal V}'(\tilde{\cal B}(u)) - {\cal V}'(\tilde b^1))$.

At the beginning of the aging regime $u \ll 1$, we obtain
${\cal B}_k(u) = b_k^1 - 4 {\cal V}''(\tilde b^1) (k^2 + \ov M )^{-2}
u^{\alpha}$,
and thus for $t' \sim t$:
\beq
\tilde{b}(t,t') - \tilde b^1 \sim
\ln^\alpha (h(t)/h(t'))
\sim
c(t_w)  (\tau/t_w)^\alpha
\eeq
where
$c(t_w) = (d \ln h(t_w)/d \ln t_w)^{\alpha}$.
If $h(t) = t^\delta$ we recover the trap-model scaling
where $c(t_w)$ is a constant \cite{Bo}. The exponent $\alpha$
is determined by:
\beq
\Gamma[1+2 \alpha] \Gamma[1+\alpha]^{-2}
= X~Y(\tilde b^1)
\eeq
For the power law model, $x Y(\tilde b^1)= (4-d) \gamma/(2 (1+\gamma))$
which gives $\alpha \to 0$ when $ \gamma \to \gamma_{cr}^{+} = 2/(2-d)$
and shows how ultrametricity in the case $\gamma < \gamma_{cr}$
is approached.

At widely separated times $u \to \infty$,
the approach to $b^o_k$ is
described by a scaling form of $k^2 u$:
\beqa
b_k(t,t') = b_k^1 \left(1 - \frac{x_F}{X} \right)
- \frac1{X k^2} \left(1 - \left(\frac{h(t')}{h(t)}\right)^{k^2/(X Z)}\right)
\nonumber
\eeqa
where $Z= \int du \phi(u)$ is a constant, finite
for $\gamma > \gamma_{cr}$.
Integrating over $k$ we also obtain the
large time separation behaviour
$\tilde{b}(t,t') \propto \ln^{1-d/2}(h(t)/h(t'))$.

{\it Aging regime for long range models.}
These models are solved by the ultrametric ansatz
$b_k(t,t') = \max( b_k(t,s), b_k(s,t') )$,
which inserted in (\ref{reqlt1}), (\ref{beqlt1})
leads \cite{Cukule}
to a single equation parametrized by
$\tilde b$
\beqa
 \label{chiresult}
\chi_k(t,t')
&=&
 \chi_k[\tilde b] =
(k^2 + \overline{M} - \overline{M}(\tilde b) )^{-1}
\eeqa
after simple manipulations as in \cite{Cule}.
We have defined $\overline M[\tilde{b}] =
- 4 \int_{\tilde{b}}^{\tilde{b}_k^1}
{\cal V}''(\tilde{b}') \tilde{X}[\tilde{b}']
d\tilde{b}'$ and used (\ref{Xeq})-(\ref{uuuf}).
Taking a derivative w.r.t. $\tilde b$, using
$\partial_{\tilde b} \chi_k[b_k] =
\tilde{X}[\tilde{b}] \partial_{\tilde b} b_k$,
dividing by $\tilde{X}[\tilde{b}]$ and
integrating over $\tilde b$, one gets:
\beq \label{solu}
b_k = b_k^1 +
\int_{\tilde b}^{\tilde b^1} d \tilde b'
\frac{ 4 {\cal V}''(\tilde b') }
{(k^2 + \overline{M} - \overline{M}(\tilde b'))^2}
\eeq

It also implies the self-consistency condition
$
1 = 4 {\cal V}''(\tilde b)
\int_k
(k^2 + \overline{M} - \overline{M}(\tilde b))^{-2}
$ which  coincides with the marginality condition for the replicon.
One also obtains
\beq \label{tildeX}
\tilde X( \tilde b) = - a_d {\cal V}'''(\tilde b) (-{\cal V}''(\tilde
b))^{2(d-3)/(4-d)}
\eeq
with $a_d=(4-d)(4 c_d)^{2/(4-d)}/2$.
This result, derived in a rather simple way,
is formally identical to the result of the statics \cite{Mepa}
and is here shown to apply directly to the
off-equilibrium dynamics.
The self-energy of the statics
$[\sigma](u)$ is thus formally
identified with $\overline{M} - \overline{M}(\tilde b)$.

One general prediction for the aging regime of manifolds
is the existence of a scaling form for $b_{r}(t,t')$
as a function of $r$ and the two-times quantity
$\tilde b(t,t')$:
\beq
b(r,t,t') = b_{r=0}(t,t') H[ r/b_{r=0}(t,t')^{\frac{1}{2 \zeta}} ]
\eeq
which holds at large scales, i.e large $r$ and
large, widely separated times. One has $H[0]=1$.
The Hartree method yields a remarkably simple analytical form.
Using \ref{solu} one finds \cite{Cukule}
$\partial_{\tilde b} b_k = - 4 {\cal V}''(\tilde b) /
(k^2 + ( - 4 {\cal V}''(\tilde b) c_d )^{\frac{2}{4-d}} )^2$,
e.g in real space in $d=3$ one has
$\partial_{\tilde b} b_r =
\exp({\cal V}''(\tilde b) r/(2 \pi) )$.

In conclusion, we described, using the Hartree approximation,
the features of aging of a manifold in a random medium.
This provides a
frame of reference for the analysis of data from
experiments and simulations on realistic systems
and will allow to determine
in each case whether such a regime
exists and up to what times.

\end{document}